\documentclass[aps,prb, twocolumn, superscriptaddress,floatfix]{revtex4-1}%

\usepackage{amsfonts}
\usepackage{amsmath}
\usepackage{amssymb}
\usepackage{graphicx}
\usepackage{epstopdf}
\usepackage{textcomp}
\usepackage[normalem]{ulem}
\usepackage{color}
\usepackage{gensymb}
\usepackage{float}

\begin{document}

\preprint{}

\title{Nodeless superconducting gap in the candidate topological superconductor Sn$_{1-x}$In$_x$Te for $x$ = 0.7}

\author{M. P. Smylie}
\email{matthew.smylie@hofstra.edu}
\affiliation{Department of Physics and Astronomy, Hofstra University, Hempstead, New York 11549}
\affiliation{Materials Science Division, Argonne National Laboratory, 9700 S. Cass Ave., Lemont, Illinois 60439}
\author{Kaya Kobayashi}
\affiliation{Research Institute for Interdisciplinary Science, Okayama University, Okayama 700-8530, Japan}
\affiliation{Department of Physics, Okayama University, Okayama 700-8530, Japan}
\author{T. Takahashi}
\affiliation{Department of Physics, Okayama University, Okayama 700-8530, Japan}
\author{C. Chaparro}
\affiliation{Borough of Manhattan Community College, The City University of New York, New York, New York 10007}
\author{A. Snezhko}
\affiliation{Materials Science Division, Argonne National Laboratory, 9700 S. Cass Ave., Lemont, Illinois 60439}
\author{W.-K. Kwok}
\affiliation{Materials Science Division, Argonne National Laboratory, 9700 S. Cass Ave., Lemont, Illinois 60439}
\author{U. Welp}
\affiliation{Materials Science Division, Argonne National Laboratory, 9700 S. Cass Ave., Lemont, Illinois 60439}


\begin{abstract}
High-pressure synthesis techniques have allowed for the growth of Sn$_{1-x}$In$_x$Te samples beyond the ambient In-saturation limit of $x$ = 0.5 (T$_c \sim$ 4.5 K).
In this study, we present measurements of the temperature dependence of the London penetration depth $\Delta\lambda(T)$ in this superconducting doped topological insulator for $x$ = 0.7, where T$_{c,onset}\approx 5$ K.
The results indicate fully gapped BCS-like behavior, ruling out odd-parity $A_{2u}$ pairing; however, odd-parity $A_{1u}$ pairing is still possible.
Critical field values measured below 1 K and other superconducting parameters are also presented. 
\end{abstract}

\date{\today}

\maketitle

\section{INTRODUCTION}
The potential applications of exploiting the non-Abelian statistics of Majorana-like quasiparticle excitations in topological superconductors for quantum computing schemes \cite{0a,0b,0c} has driven much research into finding candidate materials.
Majorana modes indicative of topological surface states have been observed \cite{1a,1b,1c} in superconductor/topological insulator interfaces, while high-resolution STM experiments revealed Majorana modes on vortex cores in several Fe-based superconductors \cite{x1a,x1b,x1b1,x1c,x1d,x1e}.
Doping investigations involving topological insulators (TIs) \cite{TIa,TIb}, where gapless surface states are protected by time-reversal symmetry, and topological crystalline insulators (TCIs) \cite{2b,2c}, where crystalline mirror symmetry protects the gapless surface states, have resulted in discovery of unique superconducting states in $M_x$Bi$_2$Se$_3$ ($M$= Cu, Sr, Nb) and Sn$_{1-x}M_x$Te ($M$= Ag, Pb, In) \cite{3a, 3b, 3c} with transition temperatures on the order of a few kelvin.
The family of doped bismuth selenides all exhibit nematic superconductivity \cite{4a} and an odd-parity pseudo-triplet state \cite{Fu1,Fu2,Fu3} with a highly anisotropic or nodal superconducting gap, which is consistent with a topological state.
The In-doped tin telluride shows possible odd-parity pairing and a full superconducting gap \cite{3d,3e,3f}, which again is consistent with a topological state \cite{3g,3g2}.

Intrinsic superconductivity below 0.3 K arising from antisite defects in SnTe and InTe have been known since the late 1960’s \cite{2e}.
Preliminary studies \cite{3} involving In-doping on the Sn site raised T$_c$ nearly an order of magnitude without significant increase in the low ($\sim$10$^{21}$ cm$^{-3}$) carrier concentration.
Following the prediction and establishment \cite{2c,2d} of SnTe as a TCI in 2012, more recent efforts \cite{3h,3i} using modified floating-zone methods have raised the onset of superconductivity in Sn$_{1-x}$In$_x$Te to 4.5 K, with an essentially linear increase in T$_c$ with increasing $x$ for $x >$ 0.04.
The topological state is still observed at least to x = 0.4 \cite{PolleyARPES}.
Above $x$ = 0.5, In was found to be no longer soluble in the FCC SnTe structure, and significant tetragonal InTe is formed instead (which is not a TCI).
Under pressure a phase transition from tetragonal to rocksalt FCC is observed \cite{3j} in InTe; recent high-pressure synthesis efforts \cite{5} have increased the solubility of In in the SnTe structure, allowing the entire doping window from $x$ = 0 to $x$ = 1 to be accessible.
FCC-phase InTe is still believed to be topologically trivial, but recent pressure-dependent Raman spectroscopy measurements \cite{Rajaji} suggest doping FCC-phase InTe may generate a topologically nontrivial state.
The superconducting transition temperature peaks near $x$ = 0.7 with an onset temperature of $\sim$5 K, then decreases with higher indium concentrations.

In this work, we report on magnetization measurements and low-temperature measurements of the London penetration depth $\lambda$ in the highly doped TCI-derived superconductor Sn$_{0.3}$In$_{0.7}$Te down to $\sim$0.46 K.
The observed temperature dependence of $\lambda$ indicates a full superconducting gap, which eliminates one of two possible candidate topological superconducting states for Sn$_{0.3}$In$_{0.7}$Te.

\section{EXPERIMENTAL METHODS}
Polycrystalline ingots of Sn$_{0.3}$In$_{0.7}$Te were grown following the method of Ref. \onlinecite{5}.
High purity powder of Te, Sn, and shots of In were weighed in stoichiometric ratio and sealed in an evacuated tube.
The mixture was heated at 850 \degree C for a day to obtain cubic SnTe and tetragonal InTe.
The pelletized powder was placed in the high pressure cell and treated at 500 \degree C under 2 GPa for 30 min.
Powder x-ray diffraction of ground samples was taken at room temperature with a Rigaku 1100 diffractometer.
Following synthesis, samples were kept in a freezer to avoid the deformation from the metastable cubic phase into the tetragonal InTe form.
Measurements were performed before the deformation takes place.

Preliminary magnetometry measurements were performed on an irregularly shaped single ingot ($\sim$5 mm x 3 mm x 2 mm) with a custom-built SQUID magnetometer with a small conventional magnet down to 1.2 K.
Further magnetization measurements were performed on a thin sliver cut from this piece with a Quantum Design MPMS dc SQUID magnetometer with a superconducting magnet down to 1.8 K.
The tunnel diode oscillator (TDO) technique \cite{39} was used on another piece cut from the bulk ($\sim$800 um x 600 x 200 um) to measure the temperature dependence of the London penetration depth $\Delta\lambda(T) = \lambda(T) - \lambda_0$, where $\lambda_0$ is the zero-temperature value, down to $\sim$460 mK in an Oxford $^3$He cold-finger cryostat with a custom resonator \cite{20} running at $\sim$14.5 MHz.

\section{RESULTS AND DISCUSSION} 
\begin{figure}
	\includegraphics[width=1\columnwidth]{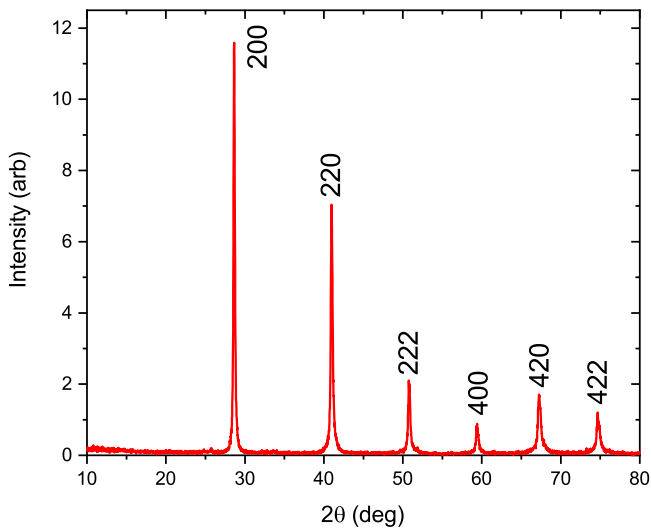}
	\caption{
		Powder x-ray diffraction patterns for Sn$_{0.3}$In$_{0.7}$Te.
		Peaks are indexed as shown, indicating rocksalt-like structure.
		No impurity peaks are observed.
	}
	\label{Fig1-XRD}
\end{figure}

Powder XRD measurements are shown in Fig.~\ref{Fig1-XRD}.
A clear rocksalt-like structure, space group $Fm\bar{3}m$, is observed; peaks are indexed as shown.
No impurity peaks from the tetragonal InTe phase (or any others) are observed.
The refined lattice constant is 6.223 $\pm$ 0.001 \AA, which is in good agreement with previous reporting \cite{5} for material with x = 0.7.
Previous reports \cite{5,Haldo,Zhang} find an almost linear relationship between lattice constant $a$ and x across the entire doping window, indicating homogenous incorporation of indium in accordance wtih Vegard's law; reported EDS measurements \cite{Haldo} show that nominal and actual doping levels are closely matched.

\begin{figure}
	\includegraphics[width=1\columnwidth]{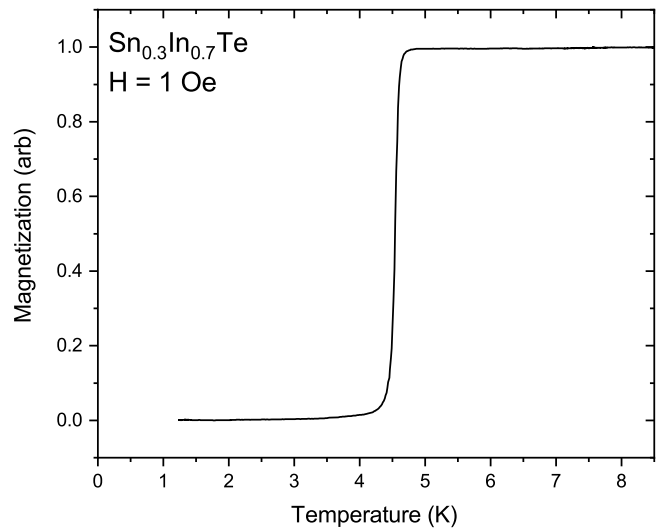}
	\caption{
		Zero field-cooled magnetization measurements with H = 1 Oe on a large ingot of polycrystalline Sn$_{0.3}$In$_{0.7}$Te.
		The transition is sharp and has an onset of T$_{c,onset} \sim$ 4.9 K.
		No secondary transitions are visible.
	}
	\label{Fig2-MT-lowT}
\end{figure}

Normalized magnetization vs temperature measurements on a single piece of Sn$_{0.3}$In$_{0.7}$Te are shown in Fig.~\ref{Fig2-MT-lowT} in an applied field of 1 Oe following cooling in zero applied field (ZFC).
The transition is sharp with an onset of T$_{c,onset}\approx$ 4.9 K and a width of $\Delta$T$_c \approx$ 0.25 K, again consistent with a doping level of x = 0.7 \cite{5}.
This transition temperature is distinctly higher than that found in samples grown without high-pressure techniques, where T$_c$ has a maximum of 4.5 K \cite{3h}.
Down to 1.25 K, no additional transitions are observed, indicating a homogenous sample with no secondary superconducting phases.

\begin{figure}
	\includegraphics[width=1\columnwidth]{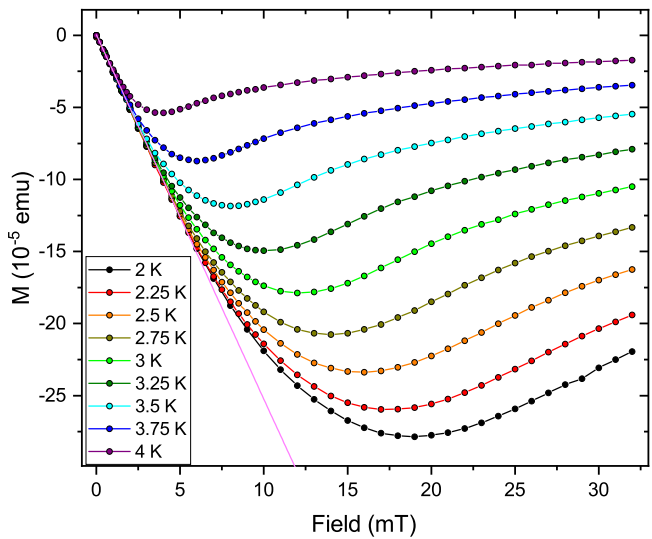}
	\caption{
		Zero-field cooled magnetization vs applied field at multiple temperatures for a thin sliver of Sn$_{0.3}$In$_{0.7}$Te.
		The lower critical field $H_{c1}$ is taken as where the magnetization first deviates from linearity (pink line).
	}
	\label{Fig3-MH}
\end{figure}

The zero-temperature London penetration depth $\lambda_0$ can be estimated from measurements of the upper and lower superconducting critical fields.
$H_{c1}$ values were deduced from low-temperature magnetization vs field measurements shown in Fig.~\ref{Fig3-MH} for a thin plate-like sample with the field applied parallel to the plate.
We take the deviation from the linear Meissner behavior in H as the value of the critical field; $H_{c1}$ vs temperature is plotted in Fig.~\ref{Fig5-phase}(a).
The error bars represent the separation in H steps (5 G) at low fields.
With a conventional parabolic temperature dependence $H_{c1} = H_{c1}(0) \left(1 - \left(T/T_c\right)^2\right)$, we extrapolate $\mu_0 H_{c1}(T=0)$ to be 4.5 mT for Sn$_{0.3}$In$_{0.7}$Te.

\begin{figure}
	\includegraphics[width=1\columnwidth]{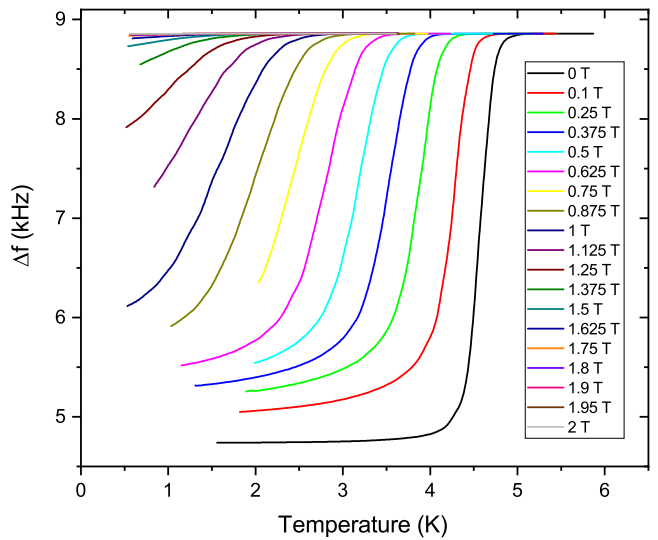}
	\caption{
		TDO frequency shift vs temperature in multiple field values for a polycrystalline sample of Sn$_{0.3}$In$_{0.7}$Te.
		The frequency shift is proportional to the magnetic susceptibility.
		With increasing field, the transition is further suppressed.
	}
	\label{Fig4-xTH}
\end{figure}

\begin{figure}
	\includegraphics[width=1\columnwidth]{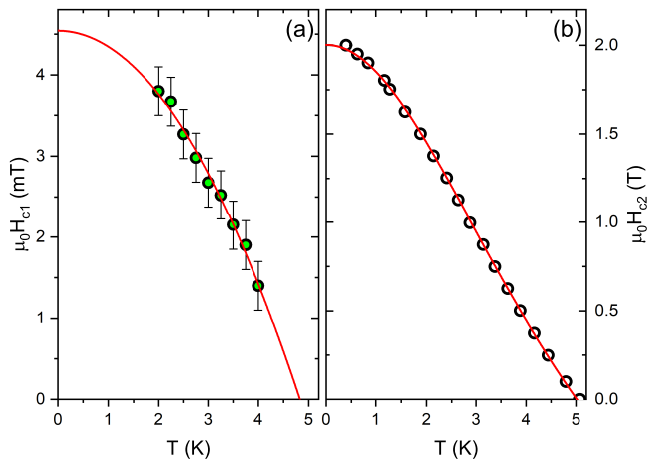}
	\caption{
		Critical field phase boundaries for Sn$_{0.3}$In$_{0.7}$Te.
		(a) Lower critical field phase boundary derived from SQUID magnetization measurements.
		(b) Upper critical field phase boundary derived from TDO susceptibility measurements.
		The lowest data point is an estimate from an incomplete transition.
	}
	\label{Fig5-phase}
\end{figure}

The shift of the TDO oscillator frequency with field and/or temperature is a measure of the degree of screening of magnetic flux in the sample which is either due to superconductivity or the normal-state skin depth \cite{39}; thus, the superconducting-normal transition is typically accompanied by a large shift in oscillator frequency allowing mapping of the temperature dependence of the upper critical field.
Such measurements of the transition in multiple field values are shown in Fig.~\ref{Fig4-xTH}.
No additional transitions are observed down to $\sim$0.46 K.
For all measurements, the sample was field-cooled in the indicated fields from above T$_c$, then data was collected during a slow warming ramp through the transition and beyond.
Defining T$_{c,onset}$ to be when the TDO resonant frequency has shifted downward by 5 Hz ($\sim$15x the noise level of a single temperature sweep) from the essentially temperature-independent normal state value yields the $H_{c2}(T)$ data shown in Fig.~\ref{Fig5-phase}(b).
This criterion emphasizes the onset of superconductivity and may overestimate $H_{c2}$ as compared to other techniques such as the resistive midpoints.
A phenomenological fit to $H_{c2}(T) = H_{c2}(0) \left( 1-t^2 \right) / \left(1+t^2 \right)$, shown in red in Fig.~\ref{Fig5-phase}(b), describes the data well, as has been observed for other superconducting doped topological insulators.
The fit extrapolates to an upper critical field at T = 0 of $\mu_0 H_{c2}(0) = 2.0$ T.
This is essentially equal to the $H_{c2}(0)$ found in Sn$_{0.55}$In$_{0.45}$Te, which nevertheless has $\sim$20\% lower T$_c$ \cite{3f}.
From our value of $H_{c2}(0)$, using the Ginzburg-Landau (GL) relation $\mu_0 H_{c2} = \Phi_0 / 2\pi\xi^2(0)$, we calculate the zero-temperature GL coherence length $\xi_0$ to be 13 nm.
Additionally, using the GL relation $H_{c1} = \Phi_0 / \left( 4 \pi \lambda^2 \right) \left( \mathrm{ln} [\lambda/\xi] + 0.5 \right)$, we estimate the zero-temperature London penetration depth $\lambda_0$ to be approximately 350 nm, consistent with observations of decreasing $\lambda$ with increasing In doping \cite{3f}.

The low-temperature penetration depth measurements were carried out via the TDO technique in the temperature range from 0.46 to 10 K.
In the TDO technique, the frequency shift $\Delta f$ of the resonator is proportional to the change of the penetration depth \cite{39}, $\Delta f(T) = G\Delta\lambda(T)$, where G is a geometrical factor which depends on the sample volume and shape as well as the resonant coil geometry.
The magnetic field inside the resonator coil is $< 5~\mu$T, assuring the sample remains fully in the Meissner state during zero-field measurement.
In the low temperature limit, the temperature dependence of $\Delta\lambda$ provides information on the superconducting gap structure; different gap structures generate different temperature dependences in $\Delta\lambda(T)$ due to the presence or absence of low-energy quasiparticles.
Conventional BCS theory for a nodeless, isotropic $s$-wave superconductor yields an exponential variation of $\Delta\lambda(T)$:
\begin{equation}
\centering
	\frac{\Delta\lambda(T)}{\lambda_0} \approx \sqrt{ \frac{\pi\Delta_0}{2T} } ~\mathrm{exp} \left(- \frac{\Delta_0}{T} \right)
\end{equation}
where $\Delta_0$ and $\lambda_0$ are the zero-temperature values of the energy gap and the penetration depth, respectively.
In a nodal superconductor, the enhanced thermal excitation of quasiparticles near the gap nodes results in a power-law variation instead, with $\Delta\lambda \sim T^n$ \cite{39,51}.
The exponent $n$ depends on the degree of electron scattering and the nature of the nodes (lines, points, etc).

\begin{figure}
	\includegraphics[width=1\columnwidth]{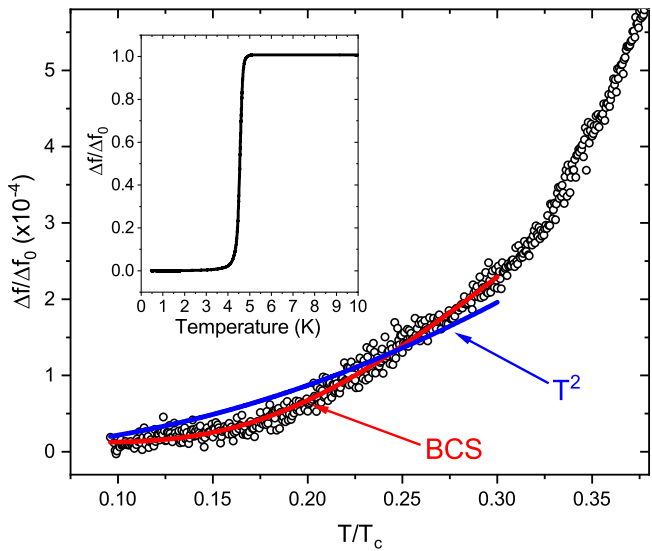}
	\caption{
		Normalized low-temperature TDO frequency shift for a thin sliver of Sn$_{0.3}$In$_{0.7}$Te with $T_{c(H=0)}$ estimated at 4.9 K.
		The BCS-like fit with a full superconducting gap (red) describes the data well; a $T^2$ fit (blue), characteristic of a gap with point nodes, is a poor fit over the same range.
		The inset shows the entire transition, with no evidence for other superconducting phases.
	}
	\label{Fig6-lambda}
\end{figure}

In Fig.~\ref{Fig6-lambda}, the relative TDO frequency shift $\Delta f / \Delta f_0$ is plotted vs reduced temperature $T/T_c$ for a thin sliver of Sn$_{0.3}$In$_{0.7}$Te cut from the large ingot used for SQUID magnetization measurements.
The results are representative of multiple measured samples.
At low temperature, a BCS-like exponential fit describes the data well albeit with a low gap ratio of $\Delta_0/T \approx 1.0$.
This value which is below the BCS $s$-wave value of 1.76, but is still consistent with a weakly anisotropic single gap or multigap superconductivity\cite{53,54,55}.
While multiple electron and hole Fermi surface sheets could support multigap superconductivity, specific heat measurements \cite{5} have not shown any evidence yet.

As no spontaneous rotational symmetry breaking is yet known to occur in the Sn$_{1-x}$In$_x$Te system, theoretical considerations \cite{3d,56} pertaining to this FCC system consider only the $A_{1g}$, $A_{1u}$ and $A_{2u}$ one-dimensional representations of the $D_{3d}$ point group as possible pairing symmetries.
The $A_{1g}$ state is the conventional, topologically trivial $s$-wave superconductor with a full superconducting gap.
The $A_{1u}$ and $A_{2u}$ are odd-parity states which are topologically nontrivial.
The $A_{1u}$ state is fully gapped and the $A_{2u}$ state will have symmetry-protected point nodes occurring at the intersection of the $L$-point-centered Fermi surfaces \cite{3g} with the $\Gamma L$-line in the FCC Brillouin zone.
The $A_{2u}$ state is not consistent with our observations of a full superconducting gap.
Our measurements cannot distinguish between the $A_{1g}$ and the $A_{1u}$ gap structures, and further investigations involving surface sensitive and/or phase sensitive techniques are required to settle this question. 

\section{CONCLUSIONS}

In summary, we have investigated the superconducting gap symmetry and critical fields of the TCI-derived superconductor Sn$_{0.3}$In$_{0.7}$Te, which is beyond the previously known In-saturation limit of $x$ = 0.5.
This doping regime, only available via high-pressure synthesis techniques, shows a higher T$_c$ than the previous ‘optimally doped’ samples with $x \approx$ 0.45, with a maximum T$_c$ at $x$ = 0.7.
We see no additional superconducting transitions down to $\sim$0.46 K.
Magnetic phase diagrams have been extended to below 1 K.
The extrapolated $H_{c2}$ at T = 0 is not increased over that found for $x$ = 0.45, but the observed decrease in $\lambda_0$ is consistent with trends from lower doping levels.
Samples with $x$ = 0.7 are found to have a full superconducting gap which is likely weakly anisotropic.
This gap structure is consistent with either the conventional, topologically-trivial $A_{1g}$ state, or the odd-parity, topologically-nontrivial $A_{1u}$ state.
We eliminate the odd-parity, nodal $A_{2u}$ state as a possibility.

\section{ACKNOWLEDGMENTS}
TDO and magnetization measurements at Argonne were supported by the U.S. Department of Energy, Office of Science, Basic Energy Sciences, Materials Sciences and Engineering Division.
Crystal synthesis was supported by Grant-in-Aid from the Ministry of Education, Culture, Sports, Science, and Technology (MEXT) under Grants No. 18K03540 and 19H01852.

\bibliographystyle{apsrev4-2}

\end{document}